\documentclass[twocolumn,floatfix,superscriptaddress,showpacs,showkeys]{revtex4}
\usepackage{graphicx}
\usepackage{dcolumn}
\usepackage{bm}
\usepackage{mathtools}
\usepackage{braket}
\usepackage{amssymb}
\usepackage{amsthm}
\usepackage{xurl}

\theoremstyle{definition}
\newtheorem{theorem}{Theorem}
\newtheorem{lemma}[theorem]{Lemma}
\newtheorem{proposition}[theorem]{Proposition}
\newtheorem{corollary}[theorem]{Corollary}

\theoremstyle{remark}

\DeclareMathOperator{\Tr}{Tr}

\begin{document}

\title{Strict Frequency Bandwidths from Causal Light Cones Enable Scalable Exact Gradient Evaluation in Quantum Optimization}
\author{Zhaoyang Song}
\affiliation{Center for Quantum Sciences and School of Physics, Northeast Normal University, Changchun 130024, China}
\author{Wei Zhang}
\affiliation{School of Physics, Harbin Institute of Technology, Harbin, Heilongjiang 150001, China}
\author{Ji Cao}
\affiliation{Department of Physics, College of Science, Yanbian University, Yanji, Jilin 133002, China}
\affiliation{Institute of Quantum Science and Technology, Yanbian University, Yanji, Jilin 133002, China}
\author{Xuexi Yi}
\email{yixx@nenu.edu.cn}
\affiliation{Center for Quantum Sciences and School of Physics, Northeast Normal University, Changchun 130024, China}
\author{Hong-Fu Wang}
\email{hfwang@ybu.edu.cn}
\affiliation{Center for Quantum Sciences and School of Physics, Northeast Normal University, Changchun 130024, China}
\affiliation{Department of Physics, College of Science, Yanbian University, Yanji, Jilin 133002, China}
\affiliation{Institute of Quantum Science and Technology, Yanbian University, Yanji, Jilin 133002, China}
\affiliation{Institute for Interdisciplinary Quantum Information Technology, Jilin Engineering Normal University, Changchun, Jilin 130052, China}
\affiliation{Jilin Engineering Laboratory for Quantum Information Technology, Changchun, Jilin 130052, China}  

\date{October 28, 2026}

\begin{abstract}
Dynamical locality is often regarded as a limitation of shallow quantum circuits, yet we show that it can reduce the number of circuit settings required for exact evaluation of gradients with respect to shared parameters. For layered ansätze generated by commuting local terms, strict Heisenberg light cones confine energy slices obtained by varying one shared layer parameter to finite Fourier bandwidths dictated by local causal geometry. This bandlimit recasts exact gradient evaluation as a local signal recovery problem and, when the relevant frequencies lie on a known frequency lattice, yields an exact multipoint parameter-shift rule based on the discrete sine transform. The quantum approximate optimization algorithm (QAOA) provides a canonical realization in which this mechanism is governed by the local structure of the graph. Specifically, when a cost angle is varied, the relevant geometry is encoded by the contact region in the interaction line graph. At fixed circuit depth on graphs of bounded degree, the number of distinct circuit settings required for the corresponding derivative collapses from \(O(m)\) to \(O(1)\). This shows that dynamical locality is not merely a constraint on correlation propagation, but also a computational resource that enables exact gradient evaluation with fewer circuit settings in quantum optimization.
\end{abstract}



\maketitle

\textit{Introduction.}---The interplay between dynamical locality~\cite{lieb1972finite,bravyi2006lieb,nachtergaele2010lieb} and optimization complexity is a central challenge in controlling large-scale quantum systems~\cite{osborne2006efficient,bharti2022noisy,cerezo2021cost,arrasmith2021effect,wang2021noise,volkoff2021large,haug2021capacity,van2021measurement,larocca2025barren,abbas2024challenges,preskill2018nisq,mcclean2018barren,bittel2021training,depalma2023limitations}. In the context of variational quantum algorithms~\cite{cerezo2021variational,kungurtsev2024iteration,mcclean2016theory,grimsley2019adapt,stokes2020quantum}, optimizing a cost function typically involves tuning the parameters of a parameterized quantum circuit~\cite{mitarai2018quantum,peruzzo2014variational,nakanishi2020sequential,benedetti2019parameterized,heyraud2023efficient,kandala2017hardware}. Although the cost function aggregates contributions from the entire system, the underlying dynamics are constrained by local gate operations~\cite{uvarov2021barren,lieb1972finite,bravyi2006lieb}. A fundamental consequence is the existence of causal light cones, which bound operator propagation in the Heisenberg picture~\cite{haah2021quantum,tran2019locality}. However, common implementations of exact gradient estimation typically rely on conservative spectral bounds induced by the full Hamiltonian~\cite{crooks2019gradients,wierichs2022general,izmaylov2021analytic,kyriienko2021generalized}. As a result, the number of circuit settings required by such methods grows with the number of interaction terms $m$, even though the response to varying a single shared layer parameter has finite frequency support governed by local causal geometry.

The parameter-shift rule~\cite{schuld2019evaluating} and its subsequent generalizations~\cite{mari2021estimating,banchi2021measuring,wierichs2022general,izmaylov2021analytic,kyriienko2021generalized,markovich2024parameter,lai2026extended} provide a general framework for exact gradient evaluation. However, for parameter-sharing architectures, even when the generalized parameter-shift rule is extended to more complex generator spectra, the relevant spectral scales remain tied to global information encoded in the full Hamiltonian unless individual generators are instead treated term by term~\cite{theis2023proper,wiersema2024here}. The high overhead of existing exact protocols stems from a structural mismatch, with gradient estimation continuing to be governed by the global spectral scale or the number of generator terms even though the single-parameter response has locally constrained frequency support. Resolving this mismatch shifts the circuit-setting cost of exact gradient evaluation away from global spectral information and onto the causal geometry.

In this Letter, we show that dynamical locality imposes precise geometric constraints on the frequency support of single-parameter energy slices, and that these constraints can be turned into an exact gradient-evaluation protocol. This resolves the mismatch identified above by showing that the frequency support of the response to varying one shared layer parameter is governed not by the global spectral scale of the full Hamiltonian but by the local interaction regions associated with the Heisenberg light cone of each cost term. For layered ansätze with commuting local generators, this causal restriction implies a strict finite Fourier bandwidth and, when the relevant frequencies lie on a known frequency lattice, enables exact gradient reconstruction via the discrete sine transform (DST). QAOA~\cite{farhi2014quantum} on bounded-degree graphs provides a concrete graph-local realization. For cost-angle slices, the bandwidth is set by the contact region in the interaction line graph, so the number of circuit settings per gradient component collapses from $O(m)$ to $O(1)$ at fixed depth. More broadly, this work suggests that causal structure is not merely a limitation to be circumvented~\cite{farhi2020typical,farhi2020worst,marwaha2021local,bravyi2020obstacles}, but a geometric organizing principle that can be directly exploited to reduce the complexity of exact gradient evaluation in quantum optimization.

\textit{Model and Geometric Framework.}---To make the causal bandwidth mechanism explicit, we specialize to a depth-$p$ QAOA ansatz on a graph $G=(V,E)$ acting on $n:=|V|$ qubits with $m:=|E|$ cost terms \cite{farhi2014quantum,zhou2020quantum,blekos2024review,hadfield2019quantum,egger2021warm,he2023alignment}.
The variational state is prepared as $\ket{\psi(\boldsymbol\gamma,\boldsymbol\beta)} = U(\boldsymbol\gamma,\boldsymbol\beta)\ket{+}^{\otimes n}$, where
\begin{equation}
\begin{aligned}
U(\boldsymbol\gamma,\boldsymbol\beta)
&= \prod_{k=1}^{p} e^{-i\beta_k B}\,e^{-i\gamma_k C}, \\
B &= \sum_{v} X_v, \qquad
C = \sum_{e} w_e P_e .
\end{aligned}
\label{eq:qaoa_ansatz}
\end{equation}
We focus on commuting 2-local Pauli cost terms, which are locally Clifford equivalent to generalized Ising forms such as MaxCut \cite{lucas2014ising}, and admit a diagonal $\pm Z_i Z_j$ representation (see Appendix~\ref{sec:light-cones}). In this basis, conjugation by $e^{-i\gamma C}$ can enlarge the support of an observable only through neighboring $Z$-type couplings.
Since all cost terms commute, one cost layer expands the support by at most one graph hop, whereas the site-wise mixer $e^{-i\beta B}$ leaves the support unchanged.
The Heisenberg evolution therefore has a strict finite-depth light cone \cite{lieb1972finite,bravyi2006lieb,haah2021quantum,tran2019locality}.

To assess scalability, we first distinguish two resources involved in gradient estimation, namely the number of distinct circuit settings \(N_{\mathrm{settings}}\) and the total shot cost \(M_{\mathrm{tot}}\). The former measures how many different circuit evaluations the exact reconstruction rule requires, whereas the latter determines the statistical precision once those settings are fixed~\cite{van2021measurement}. We work with the normalized objective \( F(\boldsymbol{\gamma},\boldsymbol{\beta}) := \langle C\rangle/\mathcal N \), with \(\mathcal N = m\) for unweighted instances and \(\mathcal N=\sum_e |w_e|\) in general. In the commuting Ising basis, a single shot returns a bounded estimator \(\hat F\in[-1,1]\), so each shifted expectation value can be estimated to fixed absolute precision with \(O(1)\) shots. We use the number of distinct circuit settings as the primary measure of system-size scaling, with the variance and optimal shot allocation of the reconstructed gradient analyzed in Appendix~\ref{sec:meas}.

\begin{figure}[t]
\centering
\includegraphics[width=\linewidth]{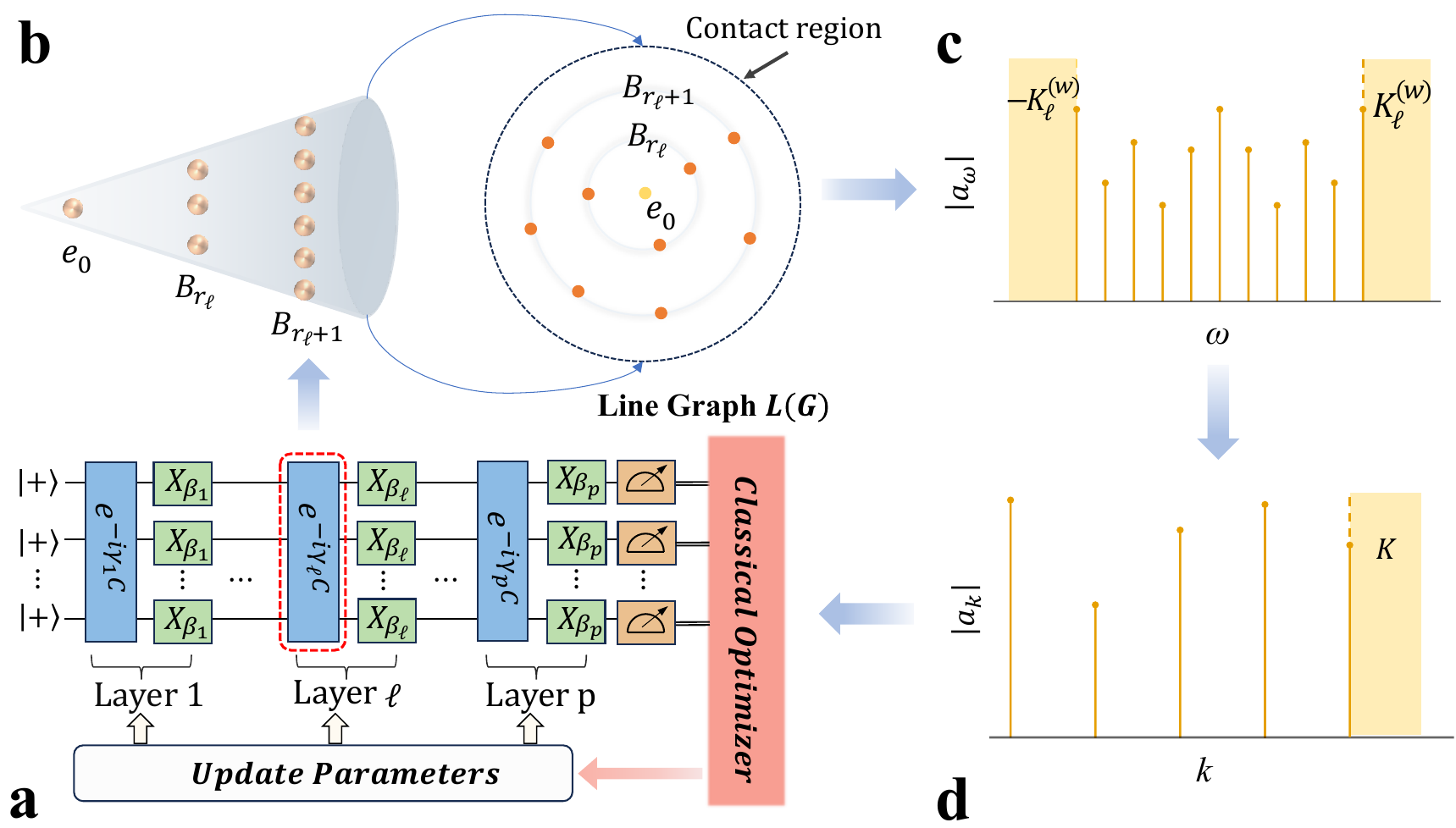}
\caption{
Causal geometry on the line graph underlying DST reconstruction of a cost-angle derivative. (a) A depth-\(p\) QAOA circuit with the target cost-angle slice \(\gamma_\ell\) highlighted. (b) Under backward Heisenberg evolution, the operator obtained from \(C_{e_0}\) is supported on the vertex set \(V(\mathcal{B}_{r_\ell}(e_0))\) induced by the corresponding ball in the line graph \(L(G)\), with \(r_\ell=p-\ell\), while only cost terms in the contact region \(\mathcal{B}_{r_\ell+1}(e_0)\) contribute to the nonzero frequencies of the slice. (c) The corresponding frequency support lies within a finite bandwidth bounded by \(K_\ell^{(w)}\). (d) For integer spectra, taking \(K=\lceil K_\ell^{(w)}\rceil\) as the DST-I cutoff yields exact reconstruction of the derivative with \(2K\) circuit settings.
}
\label{fig:geometry}
\end{figure}

For the cost-angle slice $f_\ell(\gamma_\ell):=F(\boldsymbol{\gamma},\boldsymbol{\beta})$, with all other parameters held fixed, the required number of distinct circuit settings is determined by the Fourier bandwidth of $f_\ell$ \cite{koczor2022quantum,nemkov2023fourier,casas2023multidimensional}. In the present setting, this bandwidth is fixed by a local geometric structure on the line graph $L(G)$, whose vertices represent edges of $G$ and whose adjacency is defined by shared qubits~\cite{whitney1992congruent}. Let $\mathcal{B}_r(e_0)$ denote the radius-$r$ ball centered at $e_0$ in $L(G)$, let $V(S)$ denote the set of graph vertices incident to the edges in $S\subseteq E$, and write $r_\ell:=p-\ell$ for the number of remaining problem layers. For a target term $C_{e_0}$, backward Heisenberg evolution through the remaining circuit leaves its support unchanged under the mixing layers, while each of the $r_\ell$ subsequent problem layers enlarges that support by at most one step in $L(G)$, so the evolved operator is supported on the induced vertex set $V(\mathcal{B}_{r_\ell}(e_0))$. Since the frequencies of $f_\ell$ arise from energy differences of the cost Hamiltonian $C$, only cost terms in the closed neighborhood $\mathcal{B}_{r_\ell+1}(e_0)$, which we refer to as the contact region, can contribute. The Fourier bandwidth of $f_\ell$ is therefore finite, as illustrated by the line-graph light-cone picture and the resulting cutoff in Fig.~\ref{fig:geometry}.

The spectral diameter of a local term $C_e$ is defined by \( \Delta_e:=\max_{\lambda,\lambda' \in \mathrm{spec}(C_e)}|\lambda-\lambda'|. \) For Pauli interactions, this reduces to $\Delta_e=2|w_e|$. The slice can therefore be written as 
\begin{equation}
f_\ell(\gamma_\ell)
=
\sum_{\omega \in \Omega_\ell}
\hat f_{\ell,\omega} e^{i\omega \gamma_\ell},
\end{equation}
where \( \Omega_\ell := \{\omega : \hat f_{\ell,\omega} \neq 0\} \) is the contributing frequency set.

\begin{theorem}[Causal frequency bound for cost-angle slices]\label{thm:bandwidth}
For depth-\(p\) QAOA with commuting 2-local cost terms, every nonzero frequency of the single-parameter slice \(f_\ell(\gamma_\ell)\) can be attributed to energy differences contributed by cost terms in the contact region on the line graph \(L(G)\). The physical bandwidth \(K_\ell := \max_{\omega \in \Omega_\ell} |\omega|\) satisfies
\begin{equation}
K_\ell \le K_\ell^{(w)}
:=
\max_{e_0 \in E}
\sum_{e \in \mathcal{B}_{r_\ell+1}(e_0)} \Delta_e,
\qquad
r_\ell := p-\ell.
\label{eq:bandwidth_bound}
\end{equation}
Here \(\mathcal{B}_r(e_0)\) denotes the radius-\(r\) ball centered at \(e_0\) in \(L(G)\). The effective radius \(r_\ell+1\) arises because the Heisenberg-evolved operator associated with \(C_{e_0}\) is supported on \(V(\mathcal{B}_{r_\ell}(e_0))\). Equivalently, the contact region is the closed neighborhood \(N\!\big[\mathcal{B}_{r_\ell}(e_0)\big] = \mathcal{B}_{r_\ell+1}(e_0)\). Only cost terms in this closed neighborhood can have nonvanishing commutators with the evolved observable.
\end{theorem}

Theorem~\ref{thm:bandwidth} has a simple geometric meaning: the bandwidth is determined by the contact region around $e_0$, rather than by the full Hamiltonian. For a fixed target term $C_{e_0}$, only cost terms in $\mathcal{B}_{r_\ell+1}(e_0)$ contribute, and the bound is obtained by summing their spectral diameters $\Delta_e$. In the bounded-degree MaxCut case, the size of this contact region is uniformly bounded, so the following corollary is immediate.

\begin{figure*}[t]
\centering 
\includegraphics[width=0.98\textwidth]{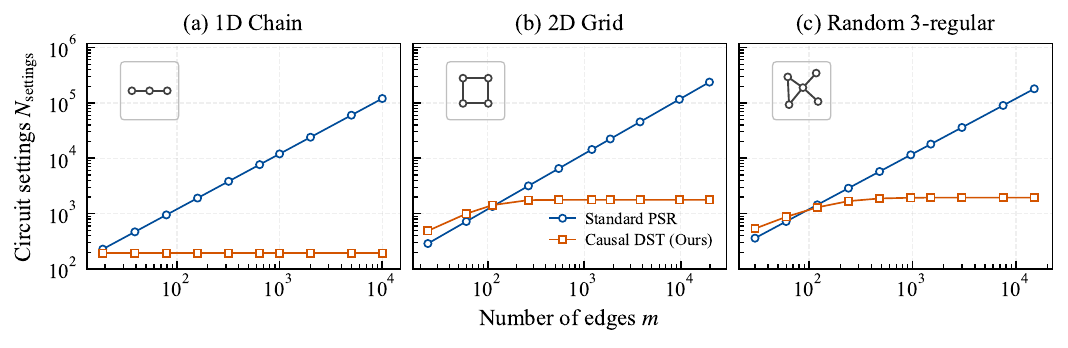}
\caption{ Scaling of the circuit-setting cost for the complete cost-angle gradient \(\nabla_{\boldsymbol\gamma}F\) at fixed depth \(p=6\) on unweighted graphs. The number of circuit settings \(N_{\mathrm{settings}}\) is shown versus the number of edges \(m\) for (a) 1D chains, (b) 2D grids, and (c) random 3-regular graphs. Blue circles denote gate-level PSR, while orange squares denote causal DST. The PSR cost grows linearly with \(m\), whereas the DST cost saturates for bounded-degree graphs, consistent with a bandwidth set by the local light cone. }
\label{fig:scaling}
\end{figure*}

\begin{corollary}[Bounded-degree MaxCut: constant bandwidth]\label{cor:maxcut}
For standard unweighted MaxCut, \( C_{\mathrm{MC}} = \sum_{e=\{i,j\}} \frac{1}{2}(I - Z_i Z_j) \). After discarding the global additive constant, each nonconstant local term is \( -\frac{1}{2} Z_i Z_j \), hence \(\Delta_e = 1\). Therefore,
\begin{equation}
K_\ell
\le
\max_{e_0 \in E}
\bigl|\mathcal{B}_{r_\ell+1}(e_0)\bigr|.
\end{equation}
Since the line-graph degree obeys
\begin{equation}
d_L := \max_{e \in E} \deg_{L(G)}(e) \le 2(d_{\max}-1),
\end{equation}
the ball size \(\bigl|\mathcal{B}_{r_\ell+1}(e_0)\bigr|\) depends only on \(d_{\max}\) and \(r_\ell\), and not on \(n\) or \(m\). Hence, for any fixed depth \(p\) and bounded \(d_{\max}\), the bandwidth \(K_\ell\) is an \(\mathcal{O}(1)\) constant independent of the system size.
\end{corollary}

A parallel locality statement holds for mixer-angle slices with all other variables fixed. Because the standard mixer \(B = \sum_v X_v\) is one-local, the relevant region is the induced vertex support of the light cone \(\mathcal{B}_{r_\ell}(e_0)\), rather than the edge contact region \(\mathcal{B}_{r_\ell+1}(e_0)\). The corresponding frequencies are thus bounded by local mixer-energy differences within this induced vertex set, which again yields an \(\mathcal{O}(1)\) bandwidth at fixed depth on bounded-degree graphs. Because the standard mixer has integer spectrum, the same exact DST reconstruction applies to \(\partial_{\beta_\ell} F\). The precise theorem and proof are given in Appendix~\ref{sec:light-cones}.

\textit{Exact Gradient Recovery via DST.}---In the integer-spectrum regime, such as unweighted MaxCut, the eigenvalues of \(C\) are integers up to an additive constant. Hence every frequency in the Fourier support of \(f_\ell(\gamma_\ell)\) is an integer \cite{nemkov2023fourier}. Combined with the bandwidth bound above, this gives \(\Omega_\ell \subset \{-K,\ldots,K\}\) for \(K := \lceil K_\ell^{(w)} \rceil\). For MaxCut, passing to the equivalent commuting Ising form changes \(C\) only by an additive constant, which affects only the \(\omega=0\) component and therefore does not affect the gradient.

For such an integer cutoff, DST-I yields an exact reconstruction formula for the derivative \cite{britanak2010discrete,nemkov2023fourier}. Using sampling shifts \(s_k = \pi k/(K+1)\) and closed-form weights \(\{\alpha_k\}\) derived in Appendix~\ref{sec:dst}, we obtain
\begin{equation}
\partial_{\gamma_\ell} f_\ell(\gamma_\ell)
=
\sum_{k=1}^{K}
\alpha_k
\Bigl[
f_\ell(\gamma_\ell+s_k)-f_\ell(\gamma_\ell-s_k)
\Bigr],
\label{eq:dst_rule}
\end{equation}
with
\begin{equation}
\alpha_k
=
\frac{1}{K+1}
\sum_{\nu=1}^{K}
\nu
\sin\!\Bigl(\frac{\pi \nu k}{K+1}\Bigr),
\qquad
k=1,\dots,K.
\end{equation}
The rule requires exactly \(2K\) distinct global circuit settings. More generally, when a rescaling factor \(c_\ell>0\) is known exactly and satisfies \(c_\ell\Omega_\ell\subset\mathbb Z\), the rescaled rule uses \(2\lceil c_\ell K_\ell^{(w)}\rceil\) settings. This condition covers integer, rational, and commensurate irrational frequency sets, whereas incommensurate frequencies fall outside the closed-form DST-I rule. The rescaling construction is given in Appendix~\ref{sec:dst}, and the corresponding mixer-angle result is given in Appendix~\ref{sec:light-cones}.

\textit{Local-Cutoff Scaling and Exactness Validation.}---For the integer-spectrum protocol, \(N_{\mathrm{settings}}=2K\) with \(K=\lceil K_\ell^{(w)}\rceil\), where the weighted cutoff \(K_\ell^{(w)}\) is set by the contact region on \(L(G)\). At fixed depth on bounded-degree graphs, the contact region has constant size. If the local spectral diameters are uniformly bounded, then \(N_{\mathrm{settings}}=\mathcal O(1)\) per gradient component. For the complete cost-angle gradient shown in Fig.~\ref{fig:scaling}, causal DST uses \(2\sum_{\ell=1}^{p}\lceil K_\ell^{(w)}\rceil\) settings, whereas gate-level PSR uses \(2pm\). At fixed \(p=6\), the DST count saturates with system size for 1D chains, 2D grids, and random 3-regular graphs, while the PSR count remains linear in \(m\). The same argument also shows what changes once \(p\) is allowed to grow with \(m\): the relevant scale is still the light-cone size, rather than the global system size \cite{mizuta2022local,akshay2021parameter}. On polynomial-growth geometries, such as \(d\)-dimensional lattices, one has \(|\mathcal{B}_R|=\Theta(R^d)\), and therefore the cutoff obeys \(K=\mathcal{O}(p^d)\). Consequently, when \(p=\mathcal O(\log m)\), the number of circuit settings grows only polylogarithmically with system size.

\begin{figure}[t]
\centering
\includegraphics[width=0.98\linewidth]{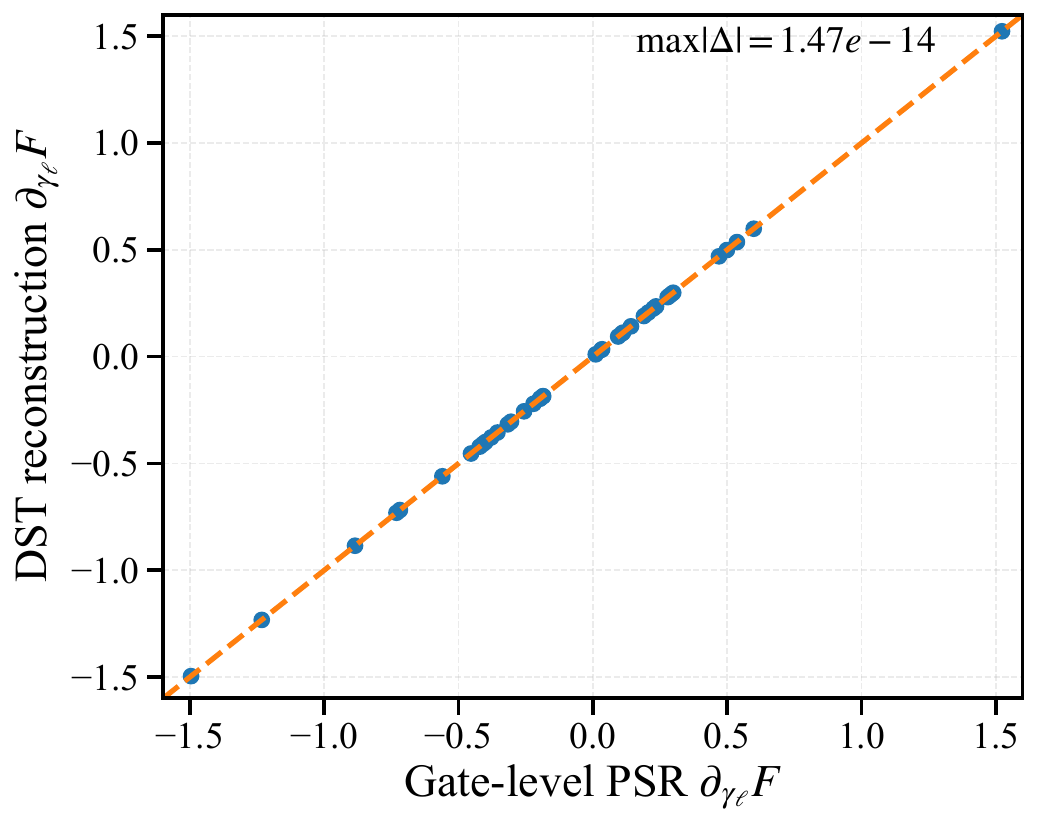}
\caption{ Exactness of the DST reconstruction for an integer-spectrum cost-angle derivative. Each point shows the DST-reconstructed value of \(\partial_{\gamma_\ell}F\) versus its gate-level PSR counterpart for one of 40 random parameter sets on a random 3-regular graph with \(n=14\) and integer weights, at layer \(\ell=2\) of a \(p=3\) circuit. The points lie on the diagonal, with a maximum absolute deviation of \(1.47\times10^{-14}\), demonstrating machine-precision agreement. }
\label{fig:equiv}
\end{figure}

This reduction in setting count comes without sacrificing exactness. As shown in Fig.~\ref{fig:equiv}, the reconstructed gradients agree with the gate-level PSR baseline to machine precision for a representative cost-angle derivative. Appendix~\ref{sec:num} further validates mixer-angle derivatives, integrated full gradients, and boundary cases with integer, rational, commensurate irrational, and incommensurate cost-angle spectral spacings.

\textit{Discussion and Outlook.}---This Letter establishes a geometric framework that connects causal structure to the complexity of exact gradient evaluation. We show that, for layered variational circuits generated by mutually commuting local generators, the circuit-setting cost of exact gradient evaluation should not be measured by the global system size, but rather by the extent of the relevant causal regions. The underlying mechanism is that the energy slice generated by varying a single shared layer parameter decomposes into contributions from individual local cost terms, each with frequency support constrained by its associated causal region. Exact gradient evaluation therefore ceases to be a problem governed by the global spectral width, and can instead be reformulated as an exact signal-reconstruction problem controlled by a local bandwidth. Following this idea, we derive an exact gradient-reconstruction rule using DST, and further show that, for bounded-degree systems at fixed depth with a uniformly bounded rescaled local bandwidth, the number of circuit settings required per gradient component can be reduced to a constant independent of system size. Our discussion of QAOA further shows that this framework is not merely abstract, but can be concretely realized in a standard variational setting.

More broadly, our work provides a geometric perspective for understanding gradient complexity in variational optimization. In the commuting local circuits considered here, this perspective yields an exact result, which in turn indicates that locality is not merely a dynamical constraint, but also a basic principle that organizes spectral structure and computational cost~\cite{stechly2024connecting,wang2025lightcone}. From this perspective, the more substantive direction is no longer a superficial extension of the present conclusions, but rather a deeper understanding of whether, in more general variational dynamics, causal constraints can still govern the effective spectral structure and support efficient gradient reconstruction~\cite{urbaneja2025exact}. Along this direction, the framework established here may serve as a starting point for analyzing the complexity of more general quantum optimization.

\vspace{\baselineskip}

\textit{Acknowledgments}---This work was supported by the Natural Science Foundation of Jilin Province under Grant No. 20240101013JC, and the National Natural Science Foundation of China under Grants No.12375020, No. 62201493, No. 62561052, and No. 12575010.

\textit{Data and Code Availability}---The data and code that support the findings of this article are openly available~\cite{code}.

\clearpage
\onecolumngrid
\appendix
\numberwithin{theorem}{section}
\numberwithin{table}{section}
\section{Heisenberg light cones and frequency band limits}
\label{sec:light-cones}

In this section we formalize the locality-to-bandwidth mechanism for commuting-local QAOA. Working in the Heisenberg picture, we construct light cones on the interaction line graph and translate the resulting geometric constraints into exact limits on the Fourier spectrum of a single-parameter energy slice~\cite{lieb1972finite,bravyi2006lieb,haah2021quantum,tran2019locality}. We first derive the cost-angle bound, then prove the parallel mixer-angle theorem, and finally collect the DST constructions needed for exact full-gradient QAOA evaluation.

\subsection{Notation and line-graph light cones}

We consider the depth-$p$ QAOA ansatz on an interaction graph $G = (V,E)$,
\begin{equation}
U(\boldsymbol\gamma,\boldsymbol\beta)
= \prod_{k=1}^p e^{-i\beta_k B}\,
e^{-i\gamma_k C},
\label{eq:s1-qaoa-ansatz}
\end{equation}
with mixing Hamiltonian $B = \sum_{v\in V} X_v$ and cost Hamiltonian
\begin{equation}
C = \sum_{e\in E} w_e P_e,
\qquad [P_e,P_{e'}]=0 \quad \forall\,e,e'\in E.
\end{equation}
Each $P_e$ is a two-qubit Pauli operator that acts nontrivially only on the qubits connected by edge $e=\{i,j\}$ and $w_e\in\mathbb{R}$.

Fix a target layer index $\ell\in\{1,\dots,p\}$ and define
\begin{equation}
U_{<\ell} = \prod_{k=1}^{\ell-1}
e^{-i\beta_k B}
e^{-i\gamma_k C},
\qquad
U_{>\ell} = \prod_{k=\ell+1}^{p}
e^{-i\beta_k B}
e^{-i\gamma_k C}.
\end{equation}
The intermediate state and the effective observable at layer $\ell$ are
\begin{equation}
\rho_\ell
= U_{<\ell}\ket{+}\!\bra{+}^{\otimes n} U_{<\ell}^\dagger,
\qquad
D_\ell
= e^{i\beta_\ell B}
U_{>\ell}^\dagger \left(\frac{C}{\mathcal N}\right) U_{>\ell}
e^{-i\beta_\ell B}.
\end{equation}
In the following support and frequency analysis we may drop the overall scalar factor $1/\mathcal{N}$, since it does not affect operator commutators, light-cone supports, or the frequency set $\Omega_\ell$.
The corresponding single-parameter energy slice is
\begin{equation}
f_\ell(\gamma_\ell)
= \Tr\!\bigl[
\rho_\ell\,
e^{i\gamma_\ell C} D_\ell
e^{-i\gamma_\ell C}
\bigr]
= \sum_{\omega\in\Omega_\ell} a_\omega e^{i\omega\gamma_\ell}.
\label{eq:s1-slice}
\end{equation}

Let $r_\ell := p-\ell$ be the number of remaining cost layers after layer $\ell$. To describe causal constraints we work on the line graph $L(G)$ whose vertices correspond to edges of $G$, with adjacency given by shared endpoints~\cite{whitney1992congruent}. For a center edge $e_0\in E$ and integer radius $r\ge 0$, define the closed ball
\begin{equation}
\mathcal{B}_r(e_0)
:= \bigl\{ e\in E
\,\big|\,
\operatorname{dist}_{L(G)}(e,e_0) \le r
\bigr\}.
\end{equation}

For each term $C_e := w_e P_e$ we introduce its local spectral diameter
\begin{equation}
\Delta_e
:= \max_{\lambda,\lambda'\in spec(C_e)}
|\lambda - \lambda'|.
\end{equation}
We define the weighted light-cone bound
\begin{equation}
K_\ell^{(w)}
:= \max_{e_0\in E}
\sum_{e'\in \mathcal{B}_{r_\ell+1}(e_0)} \Delta_{e'}.
\label{eq:s1-weighted-bound}
\end{equation}
Define the physical bandwidth by $K_\ell:=\max_{\omega\in\Omega_\ell}|\omega|$. Our goal is to prove
\begin{equation}
|\omega| \le K_\ell \le K_\ell^{(w)}
\quad
\forall\,\omega\in\Omega_\ell,
\label{eq:s1-main-ineq}
\end{equation}
which is the bandlimit inequality used throughout these appendices.

\subsection{Single-layer support growth}

We quantify how the support of a Pauli operator grows under conjugation by one QAOA layer. For an operator $O$ we denote by $\operatorname{supp}(O)$ the set of qubits on which $O$ acts nontrivially.

\begin{lemma}[Support growth]\label{lem:support-growth}
Let $C = \sum_{e\in E} w_e P_e$ with pairwise commuting two-local Pauli terms and $B = \sum_v X_v$. Let $Q$ be a Pauli operator. Then the following hold.
\begin{enumerate}
\item \emph{Mixing layer.}
\begin{equation}
	\operatorname{supp}\bigl(
	e^{i\beta B} Q e^{-i\beta B}
	\bigr)
	= \operatorname{supp}(Q).
\end{equation}
\item \emph{Problem layer.}
\begin{equation}
	\operatorname{supp}\bigl(
	e^{i\gamma C} Q e^{-i\gamma C}
	\bigr)
	\subseteq
	\operatorname{supp}(Q)
	\cup
	\bigcup_{\substack{e\in E\\ e\cap \operatorname{supp}(Q)\neq\varnothing}}
	e.
	\label{eq:s1-support-one-hop}
\end{equation}
\end{enumerate}
\end{lemma}

The first statement follows because $B$ is one-local and each factor $e^{i\beta X_v}$ acts only on qubit $v$ and does not introduce new qubits into the support. The second statement uses the commuting structure of $C$ and is most transparent after a local Clifford transformation that brings all $P_e$ to $ZZ$ form.

\subsection{Shared-vertex consistency and Clifford diagonalization}

We make the Clifford reduction explicit. The following structural property is implied by pairwise commutativity of the two-local terms.

\begin{lemma}[Shared-vertex consistency]\label{lem:shared-vertex}
Let $\{P_e\}_{e\in E}$ be a set of pairwise commuting two-local Pauli operators. If two edges $e=\{i,j\}$ and $e'=\{j,k\}$ share a vertex $j$, then the single-qubit Pauli matrices of $P_e$ and $P_{e'}$ acting on qubit $j$ are identical whenever both are nontrivial.
\end{lemma}

\begin{proof}
Write $P_e|_j = \sigma_j$ and $P_{e'}|_j = \tau_j$ with $\sigma_j,\tau_j\in\{I,X,Y,Z\}$. Since $P_e$ and $P_{e'}$ act on disjoint sets of qubits except at $j$, the commutator $[P_e,P_{e'}]$ reduces to the commutator of the single-qubit operators at $j$. Two single-qubit Pauli operators commute if and only if they are identical or one of them is the identity. Hence if both $P_e$ and $P_{e'}$ act nontrivially on $j$, their local Pauli operators on $j$ must coincide.
\end{proof}

\begin{lemma}[Bitwise Clifford diagonalization]\label{lem:clifford-diag}
Under the assumptions of Lemma~\ref{lem:shared-vertex}, there exists a tensor product of single-qubit Clifford unitaries $R = \bigotimes_{v\in V} R_v$ such that
\begin{equation}
R P_e R^\dagger = \pm Z_i Z_j
\quad
\forall\,e=\{i,j\}\in E.
\end{equation}
\end{lemma}

\begin{proof}
By Lemma~\ref{lem:shared-vertex}, for each vertex $v$ all incident edges impose a consistent Pauli direction $\sigma_v\in\{X,Y,Z\}$ whenever they act nontrivially on $v$. Choose $R_v$ as a single-qubit Clifford that maps $\sigma_v$ to $Z$. For example, $R_v = I$ if $\sigma_v = Z$ or if no edge acts on $v$, $R_v = H$ if $\sigma_v = X$ and $R_v = H S^\dagger$ if $\sigma_v = Y$, where $H$ is the Hadamard and $S$ the phase gate. Then $R = \bigotimes_v R_v$ maps every $P_e$ to a $\pm ZZ$ operator.
\end{proof}

In the rotated basis defined by $R$, the cost Hamiltonian takes the form $C' = R C R^\dagger = \sum_{e\in E} \tilde w_e Z_i Z_j$ with real coefficients $\tilde w_e$. Conjugation of a Pauli operator $Q$ by one factor $e^{i\gamma Z_i Z_j}$ can only extend its support to qubits $i$ and $j$ and does not propagate further in a single layer. Since all $Z_i Z_j$ terms commute, the net effect of $e^{i\gamma C'}$ is a sum of such local updates and cannot generate nested commutators that would enlarge the support beyond the union on the right-hand side of Eq.~\eqref{eq:s1-support-one-hop}. This completes the proof of Lemma~\ref{lem:support-growth}.

\subsection{Layerwise light cones and frequency bounds}

We combine Lemma~\ref{lem:support-growth} with the QAOA structure to obtain a light-cone bound on the support of the effective observable $D_\ell$ and a corresponding bound on the frequency content of $f_\ell(\gamma_\ell)$. Recall that
\begin{equation}
D_\ell
= e^{i\beta_\ell B} U_{>\ell}^\dagger C U_{>\ell}
e^{-i\beta_\ell B}
= \sum_{e\in E} D_\ell^{(e)},
\end{equation}
where $D_\ell^{(e)}$ denotes the contribution obtained by propagating the single term $C_e$ (up to the overall normalization $1/\mathcal N$) backwards through the future layers and through the mixing layer at depth $\ell$.

\begin{proposition}[Layerwise light cone for $D_\ell$]\label{prop:light-cone}
For each edge $e\in E$,
\begin{equation}
\operatorname{supp}\bigl(D_\ell^{(e)}\bigr)
\subseteq V\bigl(\mathcal{B}_{r_\ell}(e)\bigr),
\end{equation}
where $V(S)$ denotes the set of vertices incident to the edges in $S\subseteq E$. In particular, the support of $D_\ell$ is contained in the union of vertex sets induced by radius-$r_\ell$ balls in $L(G)$.
\end{proposition}

\begin{proof}
Each of the $r_\ell$ future cost layers can enlarge the support of a Pauli operator by at most one hop along the interaction graph, and mixing layers leave the support invariant, as in Lemma~\ref{lem:support-growth}. In terms of the line graph this corresponds to a radius-$r_\ell$ ball. Starting from the support of $C_e$ and propagating backwards through $U_{>\ell}$, the resulting operator $U_{>\ell}^\dagger C_e U_{>\ell}$ is supported on the vertex set $V(\mathcal{B}_{r_\ell}(e))$, and the mixing layer at depth $\ell$ does not enlarge this support.
\end{proof}

To connect this to frequencies, note that in the eigenbasis of $C$ we can write $C = \sum_\lambda \lambda\,\Pi_\lambda$, with projectors $\Pi_\lambda$ onto the eigenspaces of eigenvalue $\lambda$. The matrix elements of $D_\ell$ in this basis couple eigenstates with energies $\lambda$ and $\lambda'$, and the corresponding frequency appearing in $f_\ell(\gamma_\ell)$ is $\omega = \lambda - \lambda' = \sum_{e\in E} (\lambda_e - \lambda'_e)$, where $\lambda_e$ and $\lambda'_e$ are eigenvalues of $C_e$ on the two eigenstates. For each $e$ the contribution satisfies $|\lambda_e - \lambda'_e|\le \Delta_e$.

Only those terms $C_e$ that act on qubits inside the support of $D_\ell$ can contribute to such matrix elements. Terms that act entirely outside the light cone commute with $D_\ell$ and hence cannot change the eigenvalue of $C$ between the bra and ket in Eq.~\eqref{eq:s1-slice}. They contribute only to the zero-frequency component.

\begin{lemma}[Contact region]\label{lem:contact-region}
Let $S\subseteq E$ be a set of edges. The set of edges that share at least one vertex with $S$ is exactly the closed neighborhood of $S$ in the line graph $L(G)$. In particular, if $S = \mathcal{B}_{r_\ell}(e_0)$, then the set of edges that can contribute nontrivially to the frequency content generated within this light cone is contained in $\mathcal{B}_{r_\ell+1}(e_0)$.
\end{lemma}

\begin{proof}
Two edges in $G$ share a vertex if and only if the corresponding vertices are adjacent in $L(G)$. The closed neighborhood of $S$ in $L(G)$ thus consists exactly of $S$ together with all edges that share a vertex with some edge in $S$. If $S = \mathcal{B}_{r_\ell}(e_0)$, this closed neighborhood is contained in $\mathcal{B}_{r_\ell+1}(e_0)$ by definition.
\end{proof}

\begin{theorem}[Geometry-limited frequency band]\label{thm:s1-frequency}
In the commuting two-local QAOA setting, the single-parameter slice $f_\ell(\gamma_\ell)$ is a trigonometric polynomial with finite bandwidth. Its frequency set $\Omega_\ell$ satisfies
\begin{equation}
|\omega|
\le K_\ell^{(w)}
= \max_{e_0\in E}
\sum_{e'\in \mathcal{B}_{r_\ell+1}(e_0)} \Delta_{e'}
\quad
\forall\,\omega\in\Omega_\ell,
\end{equation}
where $K_\ell^{(w)}$ is the weighted light-cone bound defined in Eq.~\eqref{eq:s1-weighted-bound}.
\end{theorem}

\begin{proof}
Since the effective observable decomposes as $D_\ell=\sum_{e\in E}D_\ell^{(e)}$ and the trace is linear, we can write the energy slice as a sum of local components:
\begin{equation}
f_\ell(\gamma_\ell) = \sum_{e\in E} f_{\ell}^{(e)}(\gamma_\ell),
\quad
f_{\ell}^{(e)}(\gamma_\ell) \coloneqq \Tr\!\left[\rho_\ell\, e^{i\gamma_\ell C} D_\ell^{(e)} e^{-i\gamma_\ell C}\right].
\end{equation}
The set of frequencies $\Omega_\ell$ appearing in $f_\ell$ is contained in the union of the frequencies appearing in each component. Therefore, the frequency support satisfies:
\begin{equation}
\Omega_\ell \subseteq \bigcup_{e\in E}\Omega_\ell^{(e)}.
\end{equation}
Now, fix any frequency $\omega\in\Omega_\ell$. By the inclusion above, there must exist some edge $e$ such that $\omega\in\Omega_\ell^{(e)}$.
By Proposition~\ref{prop:light-cone}, we have
\(
\operatorname{supp}(D_\ell^{(e)})\subseteq V(\mathcal{B}_{r_\ell}(e)).
\)
By Lemma~\ref{lem:contact-region}, only interaction terms $C_{e'}$ with
\(e'\in \mathcal{B}_{r_\ell+1}(e)\)
can fail to commute with $D_\ell^{(e)}$ and thus contribute to the energy differences generating $\omega$.
Hence,
\begin{equation}
|\omega|
\le \sum_{e'\in\mathcal{B}_{r_\ell+1}(e)}\Delta_{e'}
\le \max_{e_0\in E}\sum_{e'\in\mathcal{B}_{r_\ell+1}(e_0)}\Delta_{e'}
= K_\ell^{(w)}.
\end{equation}
Since this holds for any $\omega \in \Omega_\ell$, the proof is complete.
\end{proof}

Combining Theorem~\ref{thm:s1-frequency} with the definition of $K_\ell$ yields the bandlimit inequality in Eq.~\eqref{eq:s1-main-ineq}.

\subsection{Polynomial-growth geometries}

A direct corollary of Theorem~\ref{thm:s1-frequency} gives the depth dependence on polynomial-growth graph families. Assume that \(G\) has bounded degree \(D=O(1)\), uniformly bounded local spectral diameters \(\Delta_e\le \Delta_\ast=O(1)\), and polynomial volume growth in the sense that there exist constants \(C,d>0\), independent of system size, such that
\[
|B_R^G(v)|\le C R^d
\qquad
\text{for all } v\in V(G),\; R\ge 1.
\]
Fix an edge \(e_0=(u_0,v_0)\). Every edge \(e=(x,y)\in \mathcal{B}_{r_\ell+1}(e_0)\) in the line graph has at least one endpoint in
\[
B_{r_\ell+1}^G(u_0)\cup B_{r_\ell+1}^G(v_0),
\]
and therefore
\[
|\mathcal{B}_{r_\ell+1}(e_0)|
\le
D\bigl(|B_{r_\ell+1}^G(u_0)|+|B_{r_\ell+1}^G(v_0)|\bigr)
\le
2DC\,(r_\ell+1)^d.
\]
Substituting this into the weighted light-cone bound in Eq.~\eqref{eq:s1-weighted-bound} gives
\[
K_\ell^{(w)}
\le
\Delta_\ast \max_{e_0\in E} |\mathcal{B}_{r_\ell+1}(e_0)|
=
O\!\bigl((r_\ell+1)^d\bigr).
\]
Since \(r_\ell=p-\ell\le p\), it follows that
\[
K_\ell^{(w)}=O(p^d),
\qquad
K_\ell=O(p^d).
\]
For the integer-spectrum protocol, where \(K=\lceil K_\ell^{(w)}\rceil\) and \(N_{\mathrm{settings}}=2K\), this yields
\[
N_{\mathrm{settings}}=O(p^d).
\]
In particular, when \(p=O(\log m)\), the number of circuit settings grows only polylogarithmically with system size.

\subsection{Mixer-parameter slices}

The same causal logic extends to the mixer angle at layer $\ell$. Rewriting the circuit as
\begin{equation}
U(\boldsymbol\gamma,\boldsymbol\beta)
= U_{>\ell}\,e^{-i\beta_\ell B}e^{-i\gamma_\ell C}\,U_{<\ell},
\end{equation}
we define
\begin{equation}
\widetilde\rho_\ell
:= e^{-i\gamma_\ell C}\rho_\ell e^{i\gamma_\ell C},
\qquad
\widetilde D_\ell
:= U_{>\ell}^\dagger \left(\frac{C}{\mathcal N}\right) U_{>\ell}
= \sum_{e\in E}\widetilde D_\ell^{(e)},
\end{equation}
where $\widetilde D_\ell^{(e)}$ is the contribution obtained by propagating the single term $C_e$ backwards through the future layers only. The corresponding mixer-angle slice is
\begin{equation}
f_\ell^{(\beta)}(\beta_\ell)
= \Tr\!\bigl[
\widetilde\rho_\ell\,
e^{i\beta_\ell B}\widetilde D_\ell e^{-i\beta_\ell B}
\bigr]
= \sum_{\omega\in\Omega_\ell^{(\beta)}} b_\omega e^{i\omega\beta_\ell}.
\end{equation}
For each site term $B_v:=X_v$ we introduce the local mixer spectral diameter
\begin{equation}
\Delta_v^{(B)}
:= \max_{\mu,\mu'\in spec(B_v)}
|\mu-\mu'|.
\end{equation}
We further define the mixer-angle weighted light-cone bound
\begin{equation}
K_{\ell,\beta}^{(w)}
:= \max_{e_0\in E}
\sum_{v\in V(\mathcal B_{r_\ell}(e_0))} \Delta_v^{(B)}.
\label{eq:s1-mixer-weighted-bound}
\end{equation}

\begin{proposition}[Layerwise light cone for $\widetilde D_\ell$]\label{prop:mixer-light-cone}
For each edge $e\in E$,
\begin{equation}
\operatorname{supp}\bigl(\widetilde D_\ell^{(e)}\bigr)
\subseteq V\bigl(\mathcal B_{r_\ell}(e)\bigr).
\end{equation}
\end{proposition}

\begin{proof}
Starting from the support of $C_e$, only the $r_\ell=p-\ell$ future cost layers can enlarge the support, each by at most one hop as in Lemma~\ref{lem:support-growth}, while all future mixing layers preserve the support. Therefore $U_{>\ell}^\dagger C_e U_{>\ell}$, and hence $\widetilde D_\ell^{(e)}$, is supported inside the vertex set induced by the line-graph ball $\mathcal B_{r_\ell}(e)$.
\end{proof}

\begin{theorem}[Mixer-side frequency band]\label{thm:s1-mixer-frequency}
In the same commuting two-local QAOA setting, the mixer-angle slice $f_\ell^{(\beta)}(\beta_\ell)$ is a trigonometric polynomial with finite bandwidth. Its frequency set $\Omega_\ell^{(\beta)}$ satisfies
\begin{equation}
|\omega|
\le K_{\ell,\beta}^{(w)}
= \max_{e_0\in E}
\sum_{v\in V(\mathcal B_{r_\ell}(e_0))} \Delta_v^{(B)}
\quad
\forall\,\omega\in\Omega_\ell^{(\beta)}.
\end{equation}
For the standard QAOA mixer $B=\sum_v X_v$, one has $\Delta_v^{(B)}=2$, so
\begin{equation}
K_{\ell,\beta}^{(w)}
= 2\max_{e_0\in E}\bigl|V(\mathcal B_{r_\ell}(e_0))\bigr|.
\end{equation}
\end{theorem}

\begin{proof}
Decomposing $\widetilde D_\ell=\sum_{e\in E}\widetilde D_\ell^{(e)}$ and using linearity of the trace gives
\begin{equation}
f_\ell^{(\beta)}(\beta_\ell)
= \sum_{e\in E} f_{\ell,e}^{(\beta)}(\beta_\ell),
\qquad
f_{\ell,e}^{(\beta)}(\beta_\ell)
:= \Tr\!\left[\widetilde\rho_\ell\, e^{i\beta_\ell B}\widetilde D_\ell^{(e)} e^{-i\beta_\ell B}\right].
\end{equation}
Hence $\Omega_\ell^{(\beta)}\subseteq \bigcup_{e\in E}\Omega_{\ell,e}^{(\beta)}$. Fix $\omega\in\Omega_\ell^{(\beta)}$. Then there exists some edge $e$ such that $\omega\in\Omega_{\ell,e}^{(\beta)}$.
Let $S_e:=\operatorname{supp}(\widetilde D_\ell^{(e)})$. By Proposition~\ref{prop:mixer-light-cone},
\(
S_e\subseteq V(\mathcal B_{r_\ell}(e)).
\)
Now decompose the mixer Hamiltonian as
\begin{equation}
B = B_{S_e} + B_{S_e^c},
\qquad
B_{S}:=\sum_{v\in S} X_v.
\end{equation}
Because $\widetilde D_\ell^{(e)}$ acts trivially outside $S_e$, one has
\(
[B_{S_e^c},\widetilde D_\ell^{(e)}]=0.
\)
Since $B_{S_e}$ and $B_{S_e^c}$ commute, the component slice reduces to
\begin{equation}
f_{\ell,e}^{(\beta)}(\beta_\ell)
= \Tr\!\left[\widetilde\rho_\ell\, e^{i\beta_\ell B_{S_e}}\widetilde D_\ell^{(e)} e^{-i\beta_\ell B_{S_e}}\right].
\end{equation}
Therefore every nonzero frequency appearing in $f_{\ell,e}^{(\beta)}$ is generated entirely by eigenvalue differences of $B_{S_e}$. Because $B_{S_e}$ is a sum of commuting one-local terms, its spectral diameter is bounded by
\begin{equation}
\operatorname{diam}\!\bigl(spec(B_{S_e})\bigr)
\le \sum_{v\in S_e} \Delta_v^{(B)}
\le \sum_{v\in V(\mathcal B_{r_\ell}(e))} \Delta_v^{(B)}
\le K_{\ell,\beta}^{(w)}.
\end{equation}
Hence $|\omega|\le K_{\ell,\beta}^{(w)}$ for all $\omega\in\Omega_\ell^{(\beta)}$, which proves the claim.
\end{proof}

Since the standard mixer has integer spectrum, the DST construction developed below applies directly to $f_\ell^{(\beta)}(\beta_\ell)$ after replacing the cutoff by any integer upper bound on $K_{\ell,\beta}^{(w)}$.

\section{DST-based exact gradient rules}
\label{sec:dst}

In this section we derive the closed-form weights for the DST-based gradient rules. The generalized parameter-shift system applies to any finite frequency set, whereas the closed-form DST-I construction used here additionally requires the frequencies to lie on an integer lattice, either directly or after rescaling by a factor known exactly. We start from the generalized parameter-shift system, specialize to the integer-spectrum case, develop the rescaling construction for spectra on a known frequency lattice, and compute the norm of the weight vector used in the variance analysis.

\subsection{Generalized parameter-shift system}

For a fixed layer $\ell$ we write the single-parameter slice as
\begin{equation}
f_\ell(\gamma) = \sum_{\omega\in\Omega_\ell} a_\omega e^{i\omega\gamma},
\end{equation}
where $\Omega_\ell$ is a finite frequency set. The multipoint symmetric-shift estimator has the form
\begin{equation}
g_\ell(\gamma)
= \sum_{k=1}^{K} \alpha_k
\bigl[f_\ell(\gamma + s_k) - f_\ell(\gamma - s_k)\bigr].
\end{equation}
Substituting the Fourier series of $f_\ell$ and using
\(
f_\ell(\gamma + s_k) - f_\ell(\gamma - s_k)
= 2i\sum_{\omega} a_\omega \sin(\omega s_k)e^{i\omega\gamma},
\)
we obtain
\begin{equation}
g_\ell(\gamma)
= i\sum_{\omega\in\Omega_\ell} a_\omega
\Bigl(2\sum_{k=1}^K \alpha_k \sin(\omega s_k)\Bigr)
e^{i\omega\gamma}.
\end{equation}
The condition $g_\ell(\gamma) = f_\ell'(\gamma)$ for all $\gamma$ is equivalent to the frequency-moment constraints
\begin{equation}
2\sum_{k=1}^K \alpha_k \sin(\omega s_k) = \omega,
\quad
\forall\,\omega\in\Omega_\ell\setminus\{0\}.
\label{eq:s2-frequency-constraints}
\end{equation}
The zero-frequency mode cancels in the symmetric difference and requires no constraint.

In the integer-spectrum setting relevant for MaxCut and related Ising-type instances we have $\Omega_\ell\subset\mathbb{Z}$ and, by Theorem~\ref{thm:s1-frequency},
\begin{equation}
|\omega| \le K_\ell \le K_\ell^{(w)}.
\end{equation}
It is sufficient to enforce Eq.~\eqref{eq:s2-frequency-constraints} on $1\le \omega\le K$ as long as $K$ dominates the physical bandwidth. We define
\begin{equation}
K_{int,\ell} \coloneqq \bigl\lceil K_\ell^{(w)} \bigr\rceil
\end{equation}
and set $K = K_{int,\ell}$. We thus consider the linear system
\begin{equation}
2\sum_{k=1}^K \alpha_k \sin(\omega s_k) = \omega,
\qquad \omega = 1,\dots,K.
\label{eq:s2-system-integer}
\end{equation}

\subsection{DST-I nodes and closed-form weights}

We choose the shift points as DST-I nodes,
\begin{equation}
s_k = \frac{\pi k}{K+1},
\qquad k=1,\dots,K,
\label{eq:s2-dst-nodes}
\end{equation}
and define the sine matrix $S_{\omega k} = \sin(\omega s_k)$. The discrete orthogonality relation~\cite{britanak2010discrete} gives $S^\top S = \frac{K+1}{2} I$, so the solution of Eq.~\eqref{eq:s2-system-integer} is unique and given by $\boldsymbol\alpha = \frac{1}{K+1} S^\top \boldsymbol\omega$. In components,
\begin{equation}
\alpha_k
= \frac{1}{K+1}
\sum_{\nu=1}^{K}
\nu \sin\!\Bigl(\frac{\pi \nu k}{K+1}\Bigr),
\qquad k=1,\dots,K.
\label{eq:s2-dst-alpha}
\end{equation}

\subsection{Norm of DST weights}

Using Eq.~\eqref{eq:s2-dst-alpha} and the orthogonality of $S$,
\begin{equation}
\sum_{k=1}^K \alpha_k^2
= \frac{1}{2(K+1)}
\sum_{\nu=1}^K \nu^2
= \frac{K(2K+1)}{12}.
\label{eq:s2-alpha-norm}
\end{equation}

\subsection{Known frequency lattices and rescaling}

This rescaling to the integer lattice is natural from the viewpoint of Fourier expansions in variational quantum circuits~\cite{nemkov2023fourier,casas2023multidimensional}.

Suppose a rescaling factor $c_\ell>0$ is known exactly and satisfies $c_\ell\Omega_\ell\subset\mathbb Z$. Then $|\omega|\le K_\ell^{(w)}$ implies $|\nu|\le c_\ell K_\ell^{(w)}$ for $\nu=c_\ell\omega$. Define $K_{c,\ell} \coloneqq \bigl\lceil c_\ell K_\ell^{(w)} \bigr\rceil$ and the rescaled slice $h_\ell(\theta) \coloneqq f_\ell(c_\ell \theta)$. Applying the DST construction to $h_\ell$ yields a rule for $f_\ell'(\gamma)$ of the form
\begin{equation}
f_\ell'(\gamma)
= \sum_{k=1}^{K_{c,\ell}}
\alpha_k^{(\mathrm{phys})}
\Bigl[
f_\ell\bigl(\gamma + s_k^{(\mathrm{phys})}\bigr)
- f_\ell\bigl(\gamma - s_k^{(\mathrm{phys})}\bigr)
\Bigr],
\end{equation}
with $s_k^{(\mathrm{phys})} = c_\ell \frac{\pi k}{K_{c,\ell}+1}$ and $\alpha_k^{(\mathrm{phys})} = \alpha_k/c_\ell$.

\section{Measurement model, variance analysis, and optimal allocation}
\label{sec:meas}

\subsection{Estimator and variance formulation}

We focus on the energy density objective $F(\boldsymbol{\gamma}, \boldsymbol{\beta}) := \langle C \rangle / \mathcal{N}$. We use the physical shifts and weights defined above, with $c_\ell=1$ recovering the unrescaled integer-lattice case. For notational simplicity we write the cost-angle case explicitly; the mixer-angle estimator follows identically after replacing $\gamma_\ell$ by $\beta_\ell$ and using the corresponding cutoff $K_{\ell,\beta}^{(w)}$. After diagonalizing the commuting Pauli terms, a single circuit shot yields a bitstring $z \in \{0,1\}^n$, from which a bounded estimator $\hat{F}(\gamma; z) \in [-1, 1]$ is computed. Hence the single-shot variance is bounded by a constant independent of system size.

For layer $\ell$, the DST-based gradient estimator is a linear combination of energy density estimates at $2K$ physical shift points,
\begin{equation}
\hat{g}_\ell
= \sum_{k=1}^{K} \alpha_k^{(\mathrm{phys})}
\left[ \hat{F}(\gamma_\ell + s_k^{(\mathrm{phys})}) - \hat{F}(\gamma_\ell - s_k^{(\mathrm{phys})}) \right].
\end{equation}
With $M_{k,\pm}$ shots at $\gamma_\ell \pm s_k^{(\mathrm{phys})}$ and single-shot variances $\sigma_{k,\pm}^2$, the estimator variance is
\begin{equation}
\operatorname{Var}[\hat{g}_\ell]
= \sum_{k=1}^{K} \left(\alpha_k^{(\mathrm{phys})}\right)^2
\left(
\frac{\sigma_{k,+}^2}{M_{k,+}}
+ \frac{\sigma_{k,-}^2}{M_{k,-}}
\right).
\end{equation}

\subsection{Optimal shot allocation}

The resulting resource-allocation problem is closely related to measurement-cost and adaptive-shot analyses in variational algorithms~\cite{van2021measurement,gu2021adaptive}.

Minimizing $\operatorname{Var}[\hat{g}_\ell]$ under $M_{\mathrm{tot}} = \sum_{k=1}^K (M_{k,+} + M_{k,-})$ yields
\begin{equation}
M_{k,\pm}^*
= M_{\mathrm{tot}}\,
\frac{|\alpha_k^{(\mathrm{phys})}|\sigma_{k,\pm}}{
\sum_{j=1}^K
\bigl(|\alpha_j^{(\mathrm{phys})}|\sigma_{j,+}
+|\alpha_j^{(\mathrm{phys})}|\sigma_{j,-}\bigr)
}.
\label{eq:s3-opt-alloc}
\end{equation}
The corresponding optimal variance is
\begin{equation}
\operatorname{Var}^*[\hat{g}_\ell]
= \frac{1}{M_{\mathrm{tot}}}
\left(
\sum_{k=1}^K
|\alpha_k^{(\mathrm{phys})}|\bigl(\sigma_{k,+}+\sigma_{k,-}\bigr)
\right)^2.
\label{eq:s3-opt-var}
\end{equation}

\subsection{Scaling analysis}

With $K=K_{c,\ell}=\lceil c_\ell K_\ell^{(w)}\rceil$, the physical weights obey
\[
\sum_{k=1}^{K}
\left(\alpha_k^{(\mathrm{phys})}\right)^2
= \frac{1}{c_\ell^2}\sum_{k=1}^{K}\alpha_k^2
= \frac{K(2K+1)}{12c_\ell^2}.
\]
Under a homogeneous noise model $\sigma_{k,\pm}\approx\sigma$, Cauchy--Schwarz then implies
\[
\operatorname{Var}[\hat{g}_\ell]
= \mathcal{O}\!\left(
\frac{K^3}{c_\ell^2M_{\mathrm{tot}}}
\right).
\]
Thus an RMSE $\epsilon$ can be achieved with
\begin{equation}
M_{\mathrm{tot}}
= \mathcal{O}\!\left(
\frac{K^3}{c_\ell^2\epsilon^2}
\right).
\label{eq:s3-shot-scaling}
\end{equation}

For a prescribed relative RMSE $\eta$ with respect to the exact gradient $g_\ell=f_\ell'(\gamma_\ell)\neq0$, setting $\epsilon=\eta|g_\ell|$ gives
\[
M_{\mathrm{tot}}
= \mathcal{O}\!\left(
\frac{K^3}{c_\ell^2\eta^2|g_\ell|^2}
\right).
\]


\section{Implementation details and algorithms}
\label{sec:impl}

We summarize practical implementation details for the DST-based exact gradient-estimation framework. The workflow separates an offline precomputation (light-cone analysis on the line graph) from online circuit execution (shifted evaluations and aggregation). The cost-angle workflow described below is the primary workflow; for mixer-angle gradients, the same online routine is used after replacing the contact-region edge sum by the induced-vertex bound of Theorem~\ref{thm:s1-mixer-frequency}.

\subsection{Algorithm 1: Precomputation of the bandwidth bound (BFS on the line graph)}
\textbf{Goal.} Compute the geometry-limited bandwidth upper bound
\begin{equation}
K_\ell^{(w)} = \max_{e_0\in E}\sum_{e'\in \mathcal{B}_{R}(e_0)} \Delta_{e'},\qquad R=p-\ell+1.
\end{equation}

\textbf{Algorithm 1. Precomputation of the bandwidth bound (BFS).}
\begin{itemize}
\item \textbf{Input:} Interaction graph $G=(V,E)$, local spectral diameters $\{\Delta_e\}_{e\in E}$, layer index $\ell$, total depth $p$.
\item \textbf{Step 1 (Radius):} Set the effective radius $R = p - \ell + 1$.
\item \textbf{Step 2 (Line graph):} Construct the line graph $L(G)$ whose vertices correspond to edges in $E$ and adjacency indicates shared endpoints.
\item \textbf{Step 3 (Traversal):} For each center edge $e_0 \in E$:
\begin{itemize}
\item Perform a breadth-first search (BFS) on $L(G)$ from $e_0$ up to distance $R$.
\item Record the visited ball $\mathcal{B}_R(e_0)$.
\item Compute the local sum $S_{e_0} = \sum_{e' \in \mathcal{B}_R(e_0)} \Delta_{e'}$.
\end{itemize}
\item \textbf{Output:} $K_\ell^{(w)} = \max_{e_0 \in E} S_{e_0}$.
\end{itemize}

\subsection{Algorithm 2: Exact gradient estimation via DST}
\textbf{Goal.} Evaluate $\partial_{\gamma_\ell} f_\ell(\gamma_\ell)$ exactly at the expectation-value level using $2K$ shifted circuit settings, where $K$ is an integer cutoff chosen to upper-bound the (possibly rescaled) bandwidth of the slice. The same routine applies to $\partial_{\beta_\ell} f_\ell^{(\beta)}(\beta_\ell)$ after replacing $K_\ell^{(w)}$ by $K_{\ell,\beta}^{(w)}$ and shifting $\beta_\ell$ instead of $\gamma_\ell$.

\textbf{Algorithm 2. Exact gradient estimation via DST.}
\begin{itemize}
\item \textbf{Input:} Bandwidth bound $K_\ell^{(w)}$, a rescaling factor $c_\ell>0$ known exactly and satisfying $c_\ell\Omega_\ell\subset\mathbb Z$, and total shot budget $M_{\mathrm{tot}}$.

\item \textbf{Step 1 (Cutoff):} Set the integer sampling count
\begin{equation}
K = \left\lceil c_\ell K_\ell^{(w)} \right\rceil.
\end{equation}

\item \textbf{Step 2 (Nodes and weights):} Use DST-I nodes and weights
\begin{equation}
s_k = \frac{\pi k}{K+1},\qquad
\alpha_k = \frac{1}{K+1}\sum_{\omega=1}^{K}\omega\sin\!\Bigl(\frac{\pi\omega k}{K+1}\Bigr),
\qquad k=1,\dots,K.
\end{equation}

\item \textbf{Step 3 (Physical rescaling):} If $c_\ell \neq 1$, use physical shifts and weights
\begin{equation}
s_k^{(\mathrm{phys})} = c_\ell s_k,\qquad
\alpha_k^{(\mathrm{phys})} = \alpha_k/c_\ell.
\end{equation}

\item \textbf{Step 4 (Shot allocation):} Allocate shots across the shifted values of the target parameter, e.g. $\gamma_\ell \pm s_k^{(\mathrm{phys})}$ in the cost case or $\beta_\ell \pm s_k^{(\mathrm{phys})}$ in the mixer case. A variance-efficient choice is the optimal allocation in Eq.~\eqref{eq:s3-opt-alloc}; under approximately homogeneous noise this reduces to $M_{k,\pm}\propto |\alpha_k^{(\mathrm{phys})}|$.

\item \textbf{Step 5 (Aggregation):} Evaluate the corresponding shifted estimates, e.g. $\hat{F}(\gamma_\ell \pm s_k^{(\mathrm{phys})})$, and form
\begin{equation}
\hat{g}_\ell
= \sum_{k=1}^{K} \alpha_k^{(\mathrm{phys})}
\left[\hat{F}(\gamma_\ell + s_k^{(\mathrm{phys})})-\hat{F}(\gamma_\ell - s_k^{(\mathrm{phys})})\right].
\end{equation}
\end{itemize}

\subsection{Practical notes}
\begin{itemize}
\item \textbf{Caching.} For fixed $(K,c_\ell)$, the nodes and weights can be precomputed and reused across all optimization iterations.
\item \textbf{Computing $c_\ell$.} For rational weights $w_e\in\mathbb{Q}$, it suffices to choose $c_\ell$ as a common denominator for the local level spacings $\{2w_e\}$ of the terms that can contribute within the layer-$\ell$ contact region. Concretely, writing $w_e=a_e/b_e$ in lowest terms, one safe choice is $c_\ell=\mathrm{lcm}\{b_e\}$ over the relevant edges, which guarantees $c_\ell\Omega_\ell\subset\mathbb{Z}$. More generally, if all relevant local level spacings are integer multiples of a common unit $u_\ell>0$ known exactly, one may take $c_\ell=1/u_\ell$. For the standard mixer slice, one may simply take $c_\ell=1$ because the mixer spectrum is already integral.
\item \textbf{Offline vs online cost.} The BFS on $L(G)$ is performed offline per instance (and per layer $\ell$ if $R$ varies), while the online circuit-setting count is $2K$ per gradient component.
\end{itemize}


\section{Numerical validation settings}
\label{sec:num}

This section reports additional numerical results that systematically assess the DST-I-based gradient reconstruction on finite-size instances. We focus on three aspects: the numerical accuracy of the single-parameter slice formulas at the expectation-value level, the scaling of the resulting circuit-setting counts across different graph families, and the overall performance of the assembled full gradient together with several boundary spectral regimes. We report aggregate comparisons for all cost-angle derivatives, all mixer-angle derivatives, and the full gradient, and we examine the distinctions among integer, rational, commensurate irrational, and incommensurate cost-angle spectral spacings.

\subsection{Graph families}

Our tests cover several representative bounded-degree graph families with different neighborhood-growth behavior. The first class consists of paths and cycles, whose maximum degree is at most $2$ and which realize one-dimensional polynomial growth. The second class consists of two-dimensional grid graphs, whose maximum degree is at most $4$ and which also exhibit polynomial growth, but with faster local volume expansion as a function of radius. The third class is given by random $d$-regular graphs. Although their degree remains bounded, their local neighborhoods expand more rapidly, making them well suited for probing the onset of saturation in the bandwidth bound and in the required number of circuit settings as the depth $p$ increases.

\subsection{Validation tasks and reported quantities}

Our numerical study addresses four complementary questions. First, on small instances we use exact simulation to evaluate the slice function $f_\ell(\gamma)$ and its derivative $\partial_{\gamma_\ell} f_\ell(\gamma)$ directly, and we compare the resulting exact derivative against the DST reconstruction formula derived in Appendix~\ref{sec:dst} in order to test the reconstruction accuracy at the expectation-value level. Second, for the unweighted bounded-degree graph families used in the scaling tests, where $c_\ell=1$, we examine whether the cutoff
\[
K=\lceil K_\ell^{(w)}\rceil
\]
remains bounded as the system size grows, thereby testing the size dependence of the number of circuit settings required per gradient component. Third, on representative instances we compare the assembled cost-angle gradient block, the assembled mixer-angle gradient block, and the full assembled gradient against a gate-level PSR baseline, thus validating the complete gradient-assembly pipeline. Fourth, we construct four boundary cases with integer, rational, commensurate irrational, and incommensurate cost-angle spectral spacings in order to test how the DST-I reconstruction behaves under different frequency lattice structures. 

Accordingly, we report the number of circuit settings per gradient component, $N_{\mathrm{settings}}=2K$, and the maximum absolute errors observed in the full-gradient and boundary-spectrum tests.

\subsection{Integrated gradient validation and boundary-case tests}
\label{sec:sm-integrated}

We further examine the assembled gradient at the level of the full parameter set. Specifically, we compare all cost-angle derivatives, all mixer-angle derivatives, and the complete assembled gradient against gate-level PSR. Table~\ref{tab:sm-integrated-exactness} reports the corresponding maximum absolute errors on a representative random $3$-regular instance.

\begin{table}[t]
\caption{
Integrated validation of assembled QAOA gradients on a representative random $3$-regular instance
($|V|=14$, $|E|=21$, $p=3$).
The table reports the maximum absolute errors over five random parameter sets for all cost-angle derivatives, all mixer-angle derivatives, and the full gradient relative to gate-level PSR.
}
\label{tab:sm-integrated-exactness}
\begin{tabular}{lc}
\hline\hline
Validation block & Max absolute error vs.\ gate-level PSR \\
\hline
all cost-angle derivatives  & $1.665\times 10^{-14}$ \\
all mixer-angle derivatives & $1.804\times 10^{-15}$ \\
full gradient               & $1.665\times 10^{-14}$ \\
\hline\hline
\end{tabular}
\end{table}

All three errors are at the level of numerical roundoff. This shows that, on this instance, the gradient obtained by slice-wise reconstruction and subsequent assembly agrees with the gate-level PSR reference to machine precision. In other words, the single-parameter formulas derived above remain numerically accurate not only for isolated representative components, but across the full parameter set.

We next examine boundary spectral regimes. The purpose here is not to repeat the generic accuracy checks, but to distinguish cases that satisfy the exact DST-I conditions from a diagnostic incommensurate case that falls outside those conditions. Table~\ref{tab:sm-boundary} reports the maximum absolute errors over 40 random parameter sets for cost-angle and mixer-angle slices across instances with integer, rational, commensurate irrational, and incommensurate cost-angle spectral spacings. For the incommensurate instance, we additionally report a diagnostic relaxed truncation obtained by applying the unit-lattice DST-I formula with $K=\lceil K_\ell^{(w)}\rceil$, without claiming exact reconstruction.

\begin{table}[t]
\caption{
Maximum absolute errors over 40 random parameter sets in the boundary tests on a path graph with $|V|=8$, at layer $\ell=2$ of a $p=3$ circuit. The commensurate irrational and incommensurate cost-weight classes are realized with uniform $\sqrt{2}$ weights and with unit and $\sqrt{2}$ weights in the same relevant contact region, respectively.
}
\label{tab:sm-boundary}
\centering
\begin{tabular}{lllc}
\hline\hline
Slice type & Cost-weight class & Reconstruction & Max absolute error \\
\hline
cost-angle  & integer                 & exact DST-I                    & $6.217\times 10^{-15}$ \\
cost-angle  & rational                & exact DST-I                    & $7.536\times 10^{-15}$ \\
cost-angle  & commensurate irrational & exact DST-I                    & $4.663\times 10^{-15}$ \\
cost-angle  & incommensurate          & diagnostic relaxed truncation & $8.625\times 10^{-3}$ \\
mixer-angle & integer                 & exact DST-I                    & $1.749\times 10^{-15}$ \\
mixer-angle & rational                & exact DST-I                    & $2.220\times 10^{-15}$ \\
mixer-angle & commensurate irrational & exact DST-I                    & $2.220\times 10^{-15}$ \\
mixer-angle & incommensurate          & exact DST-I                    & $3.331\times 10^{-15}$ \\
\hline\hline
\end{tabular}
\end{table}

For cost-angle slices, the integer, rational, and commensurate irrational cases all agree with gate-level PSR to numerical precision. In the commensurate irrational instance, the same irrational weight is assigned to every edge in the relevant contact region, so all local spectral spacings are integer multiples of one common unit and a known $c_\ell$ maps them to the integer lattice. In the incommensurate instance, the contact region carries two weights whose ratio is irrational, so the spacings admit no common unit. The present closed-form exactness certification therefore does not apply, and the reported value is the residual produced by the diagnostic relaxed truncation. Mixer-angle reconstruction remains exact in all four cost-weight classes because the standard mixer has integer spectrum.

Finally, as a direct numerical check of the finite-bandwidth claim, we also examine the spectral support of representative slices. We find an out-of-band energy ratio of $1.856\times 10^{-30}$ for a cost-angle slice and $1.268\times 10^{-30}$ for a mixer-angle slice. Within numerical precision, the spectral weight is therefore entirely concentrated inside the predicted finite-bandwidth window, providing further support for the geometric locality analysis developed above.


\section{Resource accounting summary}
\label{sec:resource}

We summarize the resource scalings established above.

\subsection{Circuit-setting complexity}
Given an integer cutoff \(K\) chosen to upper-bound the (possibly rescaled) bandwidth of the slice
(e.g., \(K=\lceil c_\ell K_\ell^{(w)}\rceil\) with \(c_\ell=1\) in the integer-spectrum case),
the DST rule uses exactly
\begin{equation}
N_{\mathrm{settings}} = 2K
\end{equation}
distinct global circuit settings to estimate \(\partial_{\gamma_\ell} f_\ell(\gamma_\ell)\)
exactly at the expectation-value level.

\subsection{Shot complexity under homogeneous noise}
Under approximately homogeneous single-shot variance $\sigma_{k,\pm}\approx \sigma$, Eq.~\eqref{eq:s3-shot-scaling} yields the scaling
\begin{equation}
\operatorname{Var}[\hat{g}_\ell] = \mathcal{O}\!\left(\frac{K^3}{c_\ell^2M_{\mathrm{tot}}}\right),
\qquad
M_{\mathrm{tot}} = \mathcal{O}\!\left(\frac{K^3}{c_\ell^2\epsilon^2}\right)
\end{equation}
to achieve RMSE $\epsilon$.

\subsection{Offline computation}
The offline cost is dominated by computing $K_\ell^{(w)}$ via BFS on $L(G)$ up to radius $R=p-\ell+1$. This step is instance-dependent but does not scale with the number of optimization iterations, and its output can be reused throughout training.
\clearpage
\twocolumngrid

\end{document}